Editors,                                            29th July, 1996
Physical Review Letters,
1 Research Road, Box 1000
Ridge, New York 11961-0701

Dear editors,
   
  Please find enclosed a manuscript entitled "Spontaneous vortex
phase discovered?" submitted for publication in Physical Review
Letters. We study the electromagnetic properties of the recently
discovered material $ErNi_2B_2C$ and suggests that its properties
can be understood in terms of the spontaneous vortex phase of a
system with co-existence of superconductivity and ferromagnetism.
A post-cript file of figure 1 is sent as a seperate file after
this one.

  Thanks a lot for your attention,

                                                 Tai-Kai Ng
                                                   (HKUST)

***********************************************************************

\documentstyle[aps,preprint]{revtex}

\begin{document}
\draft
\title{
 Spontaneous vortex phase discovered?
}
\author{T. K. Ng\footnote[1]{Dept. of Physics, HKUST, Kowloon, Hong Kong
} and C.M. Varma}
\address{ Bell Laboratories, Lucent Technologies, Murray Hill, NJ07974}
\date{ \today }
\maketitle
\begin{abstract}
  It is argued that a spontaneous vortex phase may exist in the recently
discovered compound $ErNi_2B_2C$ at temperature below 2.3K. The consequences of
this proposal are discussed. In particular the magnetic response of the system 
are studied both above and below 2.3K and further experiments proposed.
 \end{abstract} 


\narrowtext
   Many years ago it was proposed \cite{v1,v2,t1,t2}
that exotic phases with co-existence of superconductivity and magnetism may
occur in systems with competing superconducting and ferromagnetic
components. The analysis was based on the Free energy functional\cite{v2}
\begin{eqnarray}
\label{gl}
{\em F} & = & \int{d^3r}[{1\over2}a|\psi|^2+{1\over4}b|\psi|^4+{\hbar^2\over
2m}|(\nabla-i{2e\over\hbar{c}}\vec{A})\psi|^2 + {\vec{B}^2\over8\pi}
\nonumber \\
 &  & +  
{1\over2}\alpha|\vec{M}|^2+{1\over4}\beta|\vec{M}|^4+{1\over2}\gamma^2
|\nabla\vec{M}|^2-\vec{B}.\vec{M}],
\end{eqnarray}
where $\vec{B}=\nabla\times\vec{A}$, $\vec{M}$ is magnetization and $\psi$
is the superconducting order parameter. It was shown that a stable spiral phase
where supeconductivity co-exists with spiraling magnetization or a
spontaneous vortex phase where magnetization is more or less uniform in
the system but vortices are  generated without an external magnetic field may occur. Subsequently, the spiral phase was
discovered in $ErRh_4B_4$ and $HoMo_6S_8$ compounds\cite{e1,e2} in a narrow temperature region between a superconducting phase and a ferromagnetic phase. 

   More recently, it was discovered that competition between superconductivity
and ferromagnetism may occur in a new material $ErNi_2B_2C$. We shall show in this paper
that $ErNi_2B_2C$ is a good candidate for the spontaneous vortex phase, or
that the spontaneous vortex phase will become stable under a relatively 
weak external magnetic field. (We define the spontaneous
vortex phase in the presence of magnetic field as a state where the density
of vortices present in the superconductor is larger than that given by the external field) Consequences 
of our proposal will be studied. To begin with,
we first review some basic features of the Ginsburg-Landau free energy
functional \ (\ref{gl}) where stability criteria associated with
various plausible phases is examined. The analysis of Ref.(1) and (2) are then extended to include the effects of an external magnetic field.

  The competition between magnetism and superconductivity appears in Eq.\ (\ref{gl})
as a Meissner effect of the superconducting component towards the internal magnetic
field produced by the magnetic component $\vec{B}=4\pi\vec{M}$. 
The existence of the spiral and spontaneous vortex phases in {\em F} are
direct consequences of Meissner effect where a uniform magnetization $\vec{M}$
cannot co-exist with a uniform superconducting order parameter $\psi$. For
systems with superconducting transition temperature $T_c$ higher than the magnetic
transition temperature 
$T_m$, a spiral phase may be stable at temperature $T_s$ less than but around
$T_m$.
The wavevector of the spiral is of order $Q\sim(\lambda_0\xi_M)^{-1/2}$
where $\lambda_0$ is the penetrating depth of the superconducting component
and $\xi_M\sim\gamma^2/\alpha$ is the coherence length of the magnetic
part\cite{v2}. The Meissner effect is avoided by
having a magnetization whose average is zero on a length scale much smaller than $\lambda_0$. At lower temperature, a ferromagnetic state with
superconductivity completely destroyed is usually lower in energy because
of the higher energy gain associate with magnetization ($\sim{k}_BT_m$) 
compare with the energy gained by superconductivity ($\sim(k_BT_c)^2/E_F$).
Alternatively, a spontaneous vortex state where magnetization is uniform
and superconducting component exists in the form of a vortex state may be
more stable than ferromagnetic state because of the gain in energy from
the superconducting component. However, this state exists only if
the internal magnetic field generated by the magnetization $\vec{B}$
satisfies the inequality
\begin{equation}
 H_{c1}<B\sim4\pi{M}<H_{c2},
\end{equation}
where $H_{c1}$ and
$H_{c2}$ are the lower and upper critical magnetic fields associated
with $\psi$. The inequality expresses the fact that a nearly
uniform magnetic field can be sustained in a superconductor only when
the density of vortices is such that the average distance between
them $l$ satisfies the inequality $\xi_0<l<\lambda_0$, where
$\xi_0=\sqrt{\hbar^2/2m|a|}$ is the coherence length of the
superconducting component. The spontaneous vortex state is
favored only in systems where the saturated magnetization is
not too strong or too weak compared with $H_c/4\pi$. Magnetic
anisotropy also plays strong role in deciding the relative stability of
various states. In particular, easy-axis anisotropy always disfavors spiral
(or linear polarized) states over ferromagnetic or spontaneous vortex
states. 

  Experimentally it is found that $ErNi_2B_2C$ is superconducting below $10.5K$\cite{e3} and orders
antiferromagnetically with a fundamental incommensurate wave vector of
$(0.553a*,0,0)$ below $6.0K$\cite{e4}. The magnetic moments reside
mostly on the $Er^{3+}$ ions which has a measured magnitude 
$\sim8\mu_B$\cite{e3}. $M$ Versus $H$ measurements indicate that the compound is magnetically strongly anisotropic with the $Er$ magnetic moments essentially
along only the in plane easy axis in $(100)$ and $(010)$ directions\cite{e3}.  
The same measurements with applied field along either of the in-plane
axes indicate a series of meta-magnetic (field-induced) transitions as 
function of magnetic field at temperature around $2.3K$\cite{e3,e5}. 
In particular, once the external magnetic field is significantly larger than
$H_{c1}\sim500G$, it is found that the system has 
a {\em ferromagnetic} component. The extrapolation of
$M(H)$ data back to zero applied field gives a ferromagnetic ordered moment of
roughly $0.33\mu_B/Er$\cite{e5}.  Zero-field specific heat measurement shows
also a break in the slope of $C$ Vs. $T$ curve at $T\sim2.3K$\cite{e3,e5}.
The existence of ferromagnetic component in the system is further supported
by studies on similar compound $TbNi_2B_2C$ which does not manifest
superconductivity at $T>2K$, but has a phase transition from antiferromagnetic
order to an ordered state with a weakly ferromagnetic component of $0.5\mu_B/Tb$
for $T<8K$\cite{e6}. The magnetization versus applied field behaviour for
the two compounds are found to be very similar except at low field region
where $M$ is always positive in the $Tb$ compound (no Meissner
effect). The similarity between the two compounds
suggests that the magnetic behaviour of the two compounds
are essentially identical, in particular a similar transition to an ordered state
with weakly ferromagnetic component is also perferred in the $Er$ compound
at around $2.3K$ - except that such a transition is forbidden by 
the superconducting component. The origin of the incommensurate antiferromagnetic and the transition to the weakly ferromagnetic transition is not quite clear and probably involves both the exchange interactions and the dipolar interactions which are of similar magnitude. Because of the strong easy-axis anisotropy which forbids smooth
deviations from antiferromagnetic state, the incommensurate antiferromagnetic state probably consists of spin domain walls
separating antiferromagnetic domains, and the weakly ferromagnetic state is formed by ordering of the domain walls. The distance 
between domain walls estimated from the incommensurate wave vector 
is around $19a$, giving effective magnetic moment of around
$0.42\mu_B/Er$, which is close to the experimentally
zero-field extrapolated value of $0.33\mu_B/Er$.

   At distance scale $>>$ lattice spacing, the antiferromagnetic 
component plays a negligible role and the competition between
superconductivity and weak-ferromagnetism can be described by a
Ginsburg-Landau functional similar to Eq.\ (\ref{gl}), except that the
$M^4$ term must be modifed to account for the strong
easy-axis anisotropy in this material. The internal magnetic field
created by magnetic moment of $0.33\mu_B/Er$ is approximately
$500G\sim{H}_{c1}$, which is found to be marginal for supporting
a spontaneous vortex state. However, a relatively weak magnetic
field $\sim{H}_{c1}$ should be enough to drive the system from
the spiral state into the spontaneous vortex state. In the 
following we shall
investigate this scenerio using the GL functional \ (\ref{gl}). We
shall assume that the magnetization $\vec{M}$ lies only on the x-y plane
and shall consider external fields only in in-plane directions.
The anisotropy in in-plane directions is not included in our
analysis. We shall consider $T_c>T_m$ and shall concentrate on the
behavior of the system around $T\sim{T}_s$ which is the regime of
experimental interest. The possibility of the system 
making a second order phase transition to spiral state at
$T=T_s$ but driven into spontaneous vortex by an external magnetic
field will be studied. We shall also discuss the alternative
possibility of the system making a direct first order transition
into the spontaneous vortex state from superconducting state. 
First we consider the temperature region $T>T_m$ and study changes in behavior of the system as 
$T\rightarrow{T}_m$. In this temperature range $M$ is small
and we can neglect the $M^4$ term in the GL functional.
The qualitative behaviour of the system at this temperature range can
be most easily understood by considering the London limit where
$\psi=$ constant and  neglecting the $|\nabla\vec{M}|^2$
term in $F$. It is then easy to minimize $F$ with respect to $\vec{M}$ 
and $\vec{A}$ to obtain $\vec{M}=\vec{B}/\alpha$, and $\vec{A}=\lambda_0^2
(1-4\pi/\alpha)\nabla\times\vec{B}$. Putting $\vec{M}$ and
$\vec{A}$ back into $F$, we obtain
\begin{equation}
{\em F}\sim\int{d}^3r\left[{-a^2\over2b}+(1-{4\pi\over\alpha})
{1\over8\pi}\left(\vec{B}^2+(1-{4\pi\over\alpha})\lambda_0^2(
\nabla\times\vec{B})^2\right)\right],
\label{glondon}
\end{equation}
where $\lambda_0^2=mc^2/8\pi{e}^2|\psi|^2$ is the London penetration
depth for the 'pure' superconducting component. $\alpha$ is a decreasing
function of temperature and the magnetic transition (in the absence of
superconducting component) occurs at $\alpha(T)=4\pi$. Notice that
the presence of magnetic component reduces the overall cost in magnetic 
energy of the `pure` superconductor by a factor $(1-4\pi/\alpha)$.
It also reduces the London penetration depth from $\lambda_0$ to
$\lambda=(\sqrt{1-4\pi/\alpha})\lambda_0$. The reduction in penetration
depth implies that the effective superfluid density observed in
experiment will increase rapidly as $T\rightarrow{T}_m$. As a result the 
critical field $H_c$
goes down by the same factor $\sqrt{1-4\pi/\alpha}$. An
interesting consequence of the free energy \ (\ref{glondon}) is
that the lower critical field $H_{c1}$ is not much affected by the 
presence of the magnetic component though the penetrating depth
$\lambda$ is strongly reduced as $T\rightarrow{T}_m$. To see that we 
consider the superconducting component in the extreme type II
limit $\lambda>>\xi_0$. In this limit the energy of
creating a vortex line per unit length $\epsilon\ $can be computed using the free energy in
the London limit \ (\ref{glondon}). For usual superconductors this quantity
is given 
in the London limit by $\epsilon_0\sim(\Phi_0/4\pi\lambda_0)^2
ln(\lambda_0/\xi_0)$ where $\Phi_0$ is the magnetic flux quantum.\cite{tin}
In the present case $\lambda_0\rightarrow\lambda$ which enhances
$\epsilon$. However, $\epsilon$ is also reduced by the same
factor $(1-4\pi/\alpha)$ because of the overall reduction in
free energy \ (\ref{glondon}). As a result, $\epsilon\sim
(\Phi_0/4\pi\lambda_0)^2ln(\lambda/\xi_0)$ in present case and is only
reduced through reduction in $\lambda$ in the logarithm.
Consequently, $H_{c1}$ is also not much reduced from its clean
superconductor value in the presence of magnetic component.
It is interesting to ask what happens in the system when 
external magnetic field of order $\sim$ few $H_{c1}$ is
applied. For usual superconductors the density of vortices 
is of order $(2\pi\lambda_0^2)^{-1}$ when magnetic field is of order
$\sim$ few $H_{c1}$. The density of vortices is
consistent with magnitude of external field which supplies
magnetic flux of order $\sim$ few $\Phi_0$ in area $2\pi\lambda_0^2$. 
In the present case, for external field of order $\sim$ few $H_{c1}$,
the external flux supplied in area $2\pi\lambda^2$ is of order
$H_{c1}\times2\pi\lambda^2\sim\Phi_0\times(\lambda/\lambda_0)^2$,
which is much less than one flux quantum if $\lambda<<\lambda_0$, and
seems to imply that the density of vortices is much less than
$(2\pi\lambda^2)^{-1}$ in this case. This conclusion is in
fact incorrect because the {\em total} magnetic field `sees` by the
superconductor $\vec{B}=\vec{H}+4\pi\vec{M}$ is much larger
than $\vec{H}$ as $\lambda_0>>\lambda$. It is easy to see that $\vec{B}=
\vec{H}/(1-4\pi/\alpha)$ in our approximation, and the {\em total} magnetic
flux the superconductor sees in area $2\pi\lambda^2$ is of order
 $\sim$ few $\Phi_0$, implying that the density of vortices is of
order $(2\pi\lambda^2)^{-1}$,
as in the case of usual superconductors.

   Similar analysis as above can be made in  momentum space
when the $|\nabla\vec{M}|^2$ term is included in the GL functional.
We find that qualitative behaviour of the system is not modified.
However, the divergence in $\lambda^{-1}$ as $T\rightarrow{T}_m$ 
is removed once the
$|\nabla\vec{M}|^2$ term is included. In particular, the
London penetration depth is saturated at value of order $\lambda\sim
(\lambda_0\xi_m)^{1/2}$ as $T\rightarrow{T}_s$. At temperatures $T$ very
close to $T_{s}$, the magnetic response of the system is dominated by the
spiral instability. We find that spiraling magnetization
developes around single vortex solution as $T\rightarrow{T}_s$, with
magnitude of spiral decreasing exponentially as distance 
away from vortex core.
The decaying length of spiral magnetization goes to infinity as
$T\rightarrow{T}_s$, signaling the onset of spiral instability. 
We find also that the energy of vortex line $\epsilon$ remains
finite and is given in the London limit by 
\[
\epsilon=
\left({\Phi_0^2\over4\pi\lambda_0^2}\right)\left(ln{\lambda\over\xi}-1+
O({\xi_M\over\lambda_0})\right),\]
at precisely the spiral instability point where $\lambda=
(\lambda_0\xi_M)^{1/2}$. Notice that $\lambda^{-1}$ is of the same order
of magnitude as the spiral wave vector around the spiral instability.

  The behaviour of vortices at $T<{T}_s$ can also be studied
in the London limit. In the limit $T\rightarrow{T}_s$, the magnitude of
the spiraling magnetization is small and its effect on vortices can
be estimated perturbatively. We find that the single vortex solution is
very similar to the solution above the critical temperature, except that
the decaying length $\lambda_s$ for the `extra` spiraling magnetization 
around vortex decreases again as
$T$ decreases below $T_s$, until $\lambda_s\sim(\lambda_0\xi_M)^{1/2}$,
where the perturbative solution becomes unreliable. In particular,
the energy $\epsilon$ for single vortex line remains of order
$\Phi_0^2/(2\pi\lambda_0^2)$ through out the whole temperature range,
with no discontinuity across the spiral transition point. 

  Next we consider the situation of finite density of vortices and estimate the magnetization as a function of the external magnetic field. 
Consider
the Gibb's energy functional 
\begin{equation}
{\em G}={\em F}-\int{d}^3r{\vec{B}.\vec{H}\over4\pi},
\label{gib}
\end{equation}
where the total magnetic field $\vec{B}$ is obtained by minimizing ${\em G}$
with respect to $\vec{B}$ and $\vec{M}=(\vec{B}-\vec{H})/4\pi$. First consider the regime $T>T_m$ and the
London limit using Eq.\ (\ref{glondon}). Let the applied
field strength of order $\sim$ few $H_{c1}$. The total
magnetic field $B$ can be obtained easily by comparing the present
expression for Gibb's energy with Gibb's energy for usual
superconductors\cite{tin}. We obtain
\begin{equation}
\label{mh1}
H\sim(1-{4\pi\over\alpha})B+H_{c1}{ln(H^0_{c2}/B)\over{ln}(\lambda
/\xi_0)},
\end{equation}
where $H^0_{c2}\sim\Phi_0/(2\pi\xi_0^2)$ and $M\sim{B}/\alpha-
H_{c1}[ln(H^0_{c2}/B)/ln(\lambda/\xi_0)]/4\pi$.  Notice that
the response of magnetic component $M$ to magnetic field $B$
is almost identical to response of the `pure` magnetic 
system to $B$ except the correction term
$\sim-H_{c1}$ coming from Meissner effect. 
In particular, for small enough value of $\alpha$, the magnetization
may become positive in this range of magnetic field. The value of
$H_{c2}$ where superconductivity is completely destroyed can also
be obtained easily by equating $B={H_{c2}}/(1-4\pi/\alpha)\sim{H}^0_{c2}$,
obtaining $H_{c2}\sim(1-4\pi/\alpha)H^0_{c2}$, indicating that
$H_{c2}$ is reduced in the presence of the magnetic component.
 
 The magnetization curve in temperature range $T>T_m$ thus has 
the following qualitative feature:
(1)$M=-H/4\pi$ for $H<H_{c1}$, where the value of
$H_{c1}$ does not depend too strongly on temperature, in particular
there is no singular behaviour around the spiral transition. (2)Magnetization
starts to increase at $H\sim{H}_{c1}$. For $H\sim$ few 
$H_{c1}$, the magnetic component already responds to the
external magnetic field more or less as if there is no 
superconducting component in the system. (3) the magnetization
continues to increase until at $H=H_{c2}$ where superconductivity
is destroyed. Notice that $M$ may become positive already at
magnetic field strength $H\sim$ few $H_{c1}$.

   Next  consider the magnetization curve in the spiral phase. 
We find that the magnetization curves above
and below the spiral transition are qualitatively
similar. In particular, the spiral state disappears and is
replaced by the spontaneous vortex phase in external magnetic field of
the order of several times $H_{c1}$. The argument is based on the observation
that at regime of temperature $T\leq{T}_s$, the vortex solution
is not much affected by spiral component. In particular, the value of
$H_{c1}$ stays more or less the same above and below the
spiral transition temperature. As external magnetic field is of order
$\sim$ few $H_{c1}$, the distance between vortices will be of order
$\sim\lambda\sim(\lambda_0\xi_M)^{1/2}$. 
However, this is of the same order as the period of the
spiral state. When the vortex distance is comparable
with period of spiral, the spiral state losses its meaning.
Thus we expect that at this magnetic field range, the spiral state
will smoothly crossover to the spontaneous vortex state
where the magnetic component of the system responses to
external magnetic field more or less independent of the superconducting
component as in the high temperature phase. In particular, the
magnetic response will be similar to that of a `pure` ferromagnet at
temperature $T<T_m$, in agreement with  
what is observed in $ErNi_2B_2C$ compound. The properties of the
spontaneous vortex phase can be studied by writing $\vec{M}=
\vec{M_0}+\vec{M'}$, where $M_0^2\sim(4\pi-\alpha)/\beta$ is the
spontaneous magnetization of the `pure` ferromagnetic component
at temperature $T<T_m$ and the GL functional can be expanded to second
order in $M'$. Neglecting the $|\nabla\vec{M}|^2$ term as before,
we find $\vec{M'}=(\vec{B}-4\pi\vec{M_0})/(12\pi-2\alpha)$, and 
the effective GL functional in the London limit in terms of
$\vec{B}$ and $\vec{H}$ fields has the same form as \ (\ref{glondon}),
except that the total magnetic field $B$ is coupled to an effective
external magnetic field $H_{eff}=4\pi{M}_0+\eta{H}$, where $\eta=
((6\pi-\alpha)/(4\pi-\alpha))$. The effective London penetration
depth is $\lambda=\eta^{-1}\lambda_0$ and the criteria for a stable
spontaneous vortex state is given by
\begin{equation}
\label{vs}
H_{c2}>H_{eff}=4\pi{M}_0+\eta{H}>>H_{c1}.
\end{equation}
Notice that $\lambda$ increases again as temperature lowers.
At very low temperature, $\lambda\rightarrow\lambda_0$. In fig.1
we show the ratio of total magnetic field to external field $B/H$
as a function of external field $H$ for $H>H_{c1}$ at several 
different values of temperatures computed using Eq.\ (\ref{mh1}),
with corresponding equation for $T<T_m$. 
We have choosen $H_{c2}/H_{c1}
=25.0$, $\xi_M/\xi_0=0.5$, $\alpha(T)=4\pi+60\pi(T-T_m)$ and
with saturated magnetization $4\pi{M}=2.0H_{c1}$ at zero
temperature in generating the figure. It is clear that
the zero field extrapolation of the curve at $T=0.5T_m$
indicates existence of ferromagnetic component in the system.
Notice that $B/H$ measures the total density of vortices
in the system and is $>1$ for $T<T_m$. 
The density of vortices for the 'pure' superconductor 
is close to the curve with $T=3.0T_m$. The difference
arriving from the ferromagnetic component is huge at $T\leq{T}_m$,
as can be seen from the figure.

   Lastly we want to make a few comments on the properties of the
spontaneous vortex phase, in particular in the limit when the
saturated magnetic moment is large enough and magnetic anisotropy
is strong enough so that a direct first order transition from
superconducting phase into spontaneous vortex phase occurs in the
absence of external magnetic field. In this case, the effective
magnetic field the superconductor sees is always larger than
$H_{c1}$ and there will be no Meissner effect associated with
additional external magnetic field applied on the system, i.e.
the effective $H_{c1}$ of the system is zero and
superconductivity `appears` only when vortices are pinned to
impurity sites in the system. Notice that Meissner effect exists
in the spiral phase where $H_{c1}>0$. Thus measurement
of the Meissner effect (for example, by {\em SQUID}) will 
distinguish the spiral and spontaneous vortex phase unambigiously.
Experimentally, it seems that
Meissner effect are observed in the $ErNi_2B_2c$ compound in
the $M$ vs $H$ experiment. However, the experiment
is performed in zero-field cooled environment\cite{e3,e5} indicating that
the result may not reflect the true equilibrium thermodynamic
state of the system. Thus the possibility of a zero-field spontaneous
vortex phase existing in the compound can not be ruled out. Direct observation of the spontaneous vortex phase by imaging techniques is suggested.

   We wish to acknowledge very useful discussions with Peter Gammel
and U. Yaron.

\begin{figure}
\caption{${B/H}$ vs. $H/H_{c1}$ at $H>H_{c1}$ for several values of temperatures 
computed approximately in GL theory. The contribution from ferromagnetic component is
clear from the figure.}
\end{figure}

\begin{references}
\bibitem{v1} H.S. Greenside, E.I. Blount and C.M. Varma, \prl{\bf 46}, 49 (1981)
\bibitem{v2} E.I. Blount and C.M. Varma, \prl{\bf 42}, 1079 (1979).
\bibitem{t1} M. Tachiki {\em et.al.}, Sol. State Comm.{\bf 31}, 927 (1979);
     {\em ibid}{\bf 34}, 19 (1980).
\bibitem{t2} C.G. Kuper, M. Revzen and A. Ron, \prl{\bf 44}, 1545 (1980).
\bibitem{e1} D.C. Moncton {\em et.al.}, \prl{\bf 45},2060 (1980).
\bibitem{e2} J.W. Lynn {\em et.al.}, \prl{\bf 46},368 (1981).
\bibitem{e3}B.K. Cho {\em et.al.}, \prb{\bf 52}, 3684 (1995).
\bibitem{e4}J. Zarestky {\rm et. al.}, \prb{\bf 51}, 678 (1995); S.K. Sinha
 {\em et. al.}, \prb{\bf 51}, 681 (1995).
\bibitem{e5} P.C. Canfield, S.L. Bud'ko, and B.K. Cho, preprint.
\bibitem{e6}B.K. Cho {\em et.al.}, \prb{\bf 53}, 8499 (1996); P. Dervenagas {\em et.al.},
 \prb{\bf 53}, 8506 (1996).
\bibitem{tin} see for example, M. Tinkham, 
{\em Introduction to superconductivity}, (McGraw Hill
 1975).
\end{references}
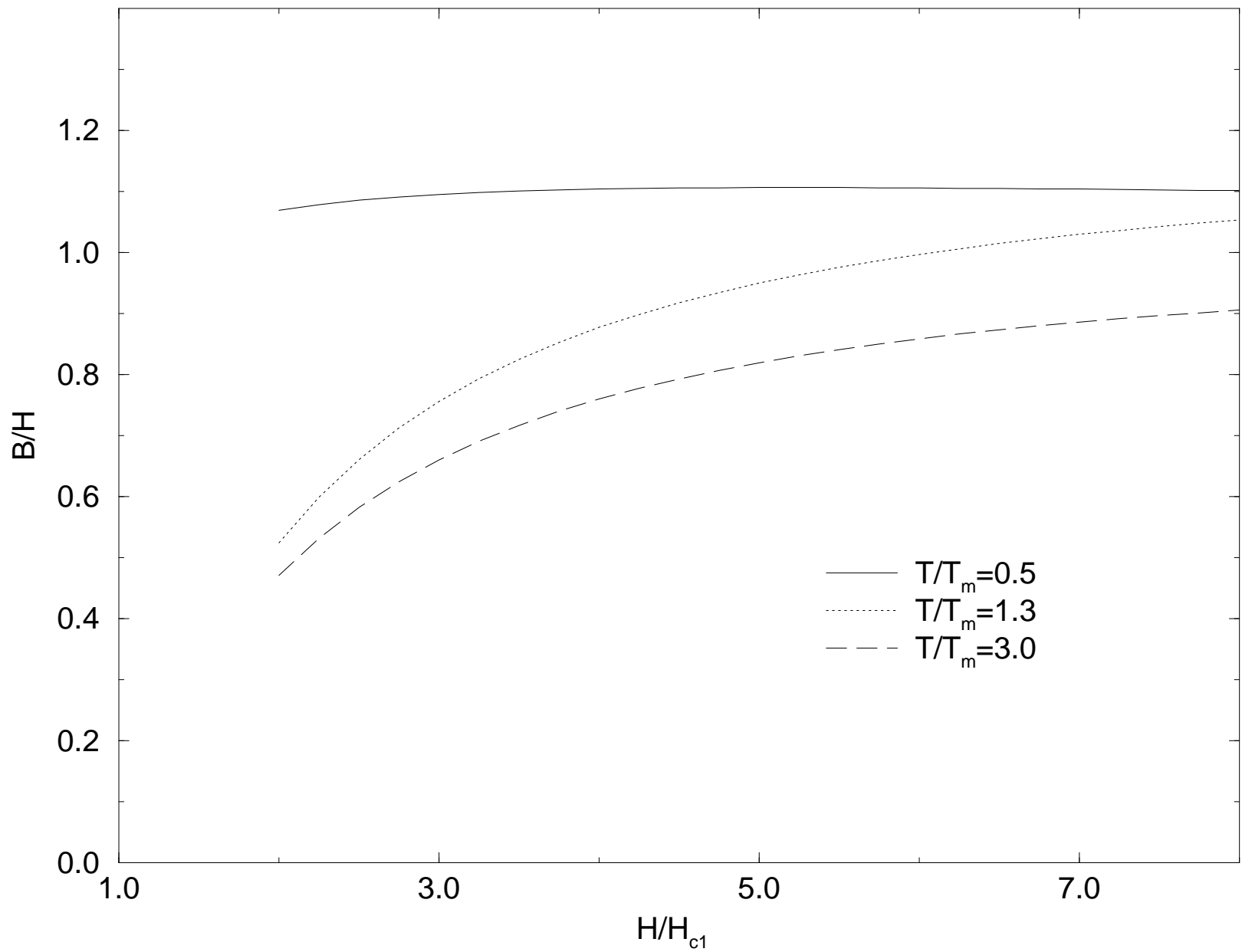
\end{document}